\documentclass[twocolumn,amsmath,amssymb,10pt,superscriptaddress,a5paper,letterpaper,fleqn,dvipsnames]{revtex4-2}

\usepackage{amssymb}
\usepackage{epsfig}
\usepackage{graphicx}
\usepackage{dcolumn}
\usepackage{array}
\usepackage{bm}
\usepackage{fancyhdr} 
\usepackage{longtable}
\usepackage{multirow}
\usepackage{float}
\usepackage{afterpage}  
\usepackage{color}
\usepackage{siunitx}
\usepackage{xcolor}
\usepackage[colorlinks=true,
    allcolors=BlueViolet,
    ]{hyperref}

\setlongtables

\def\etal{\emph{et al.}\,}

\newcommand{\RhoResult}{0.9684}
\newcommand{\RhoResultErr}{0.0018}
\newcommand{\RhoResultFull}{\mbox{$\RhoResult \pm  \RhoResultErr $}}

\newcommand{\RhoResultErrStat}{0.0013}
\newcommand{\RhoResultErrSyst}{0.0013}
\newcommand{\RhoResultSepErrs}{\mbox{$\RhoResult \stackrel{\mathrm{stat}}{\pm} \RhoResultErrStat \stackrel{\mathrm{syst}}{\pm} \RhoResultErrSyst  $}}

\newcommand{\ECStarResultPercent}{50.08}
\newcommand{\ECStarResultPercentErr}{0.06}
\newcommand{\ECStarResultPercentRelError}{0.12}

\newcommand{\ECStarResultPercentFullConciseALT}{50.08(6)}
\newcommand{\ECStarResultPercentFull}{\mbox{$(\ECStarResultPercent \pm  \ECStarResultPercentErr )\% $}}

\newcommand{\ECResultPercent}{48.50}
\newcommand{\ECResultPercentErr}{0.06}

\newcommand{\ECResultPercentFullConciseALT}{48.50(6)}
\newcommand{\ECResultPercentFull}{\mbox{$(\ECResultPercent \pm  \ECResultPercentErr )\% $}}

\newcommand{\LNHB}{Laboratoire National Henri Becquerel}
\newcommand{\NNDC}{National Nuclear Data Center}
\newcommand{\EC}{\mbox{EC$^0$}}
\newcommand{\ECs}{\mbox{EC$^*$}}
\newcommand{\betap}{\mbox{$\beta^+$}}

\newcommand{\IEC}{\mbox{$I_{\text{EC}^0}$}}
\newcommand{\IECs}{\mbox{$I_{\text{EC}^*}$}}
\newcommand{\Ibetap}{\mbox{$I_{\beta^+}$}}
\newcommand{\Zn}{\mbox{$^{65}$Zn}}
\newcommand{\Cu}{\mbox{$^{65}$Cu}}
\newcommand{\Mn}{\mbox{$^{54}$Mn}}
\newcommand{\K}{\mbox{$^{40}$K}}

\newcommand{\keV}[1]{\SI{#1}{\kilo\electronvolt}}
\newcommand{\eV}[1]{\SI{#1}{\electronvolt}}
\newcommand{\IkeV}[1]{\mbox{$\mathrm{I}(\keV{#1})$}}

\newcommand{\Kalpha}{K$_\alpha$}
\newcommand{\Kbeta}{K$_\beta$}
\newcommand{\mus}[1]{\mbox{\SI{#1}{\micro\second}}}

\newcolumntype{.}{ D{.}{.}{-1} }
\newcolumntype{x}{ D{,}{\times}{-1} }
\newcolumntype{u}{ D{,}{\pm}{-1} }
\newcolumntype{g}{ D{,}{  }{-1} }

\def\app#1#2{%
  \mathrel{%
    \setbox0=\hbox{$#1\sim$}%
    \setbox2=\hbox{%
      \rlap{\hbox{$#1\propto$}}%
      \lower1.1\ht0\box0%
    }%
    \raise0.25\ht2\box2%
  }%
}

\newcommand{\amend}[1]{\textcolor{black}{#1}}

\graphicspath{figs/}

\begin{document}
\setcounter{page}{1}

\title{Precision measurement of \texorpdfstring{\Zn}{65Zn} electron-capture decays with the KDK coincidence setup}

\newcommand{\AddOakRidgePhys}{Physics Division, Oak Ridge National Laboratory, Oak Ridge, Tennessee 37831, USA}
\newcommand{\AddOakRidgeJINPA}{Joint Institute for Nuclear Physics and Application, Oak Ridge National Laboratory, Oak Ridge, Tennessee 37831, USA}

\author{L.~Hariasz}
\author{P.C.F.~Di Stefano}\email[Corresponding author: ]{distefan@queensu.ca}
\author{M.~Stukel}
\affiliation{Department of Physics, Engineering Physics \& Astronomy, Queen's University, Kingston, Ontario K7L 3N6, Canada}
\author{B.C.~Rasco}
\author{K.P.~Rykaczewski}
\affiliation{\AddOakRidgePhys}
\author{N.T.~Brewer}
\affiliation{\AddOakRidgePhys}
\affiliation{\AddOakRidgeJINPA}
\author{R.K.~Grzywacz}
\affiliation{\AddOakRidgePhys}
\affiliation{\AddOakRidgeJINPA}
\affiliation{Department of Physics and Astronomy, University of Tennessee, Knoxville, Tennessee 37996, USA}
\author{E.D.~Lukosi}
\affiliation{Department of Nuclear Engineering, University of Tennessee, Knoxville, Tennessee 37996, USA}
\affiliation{Joint Institute for Advanced Materials, University of Tennessee, Knoxville, Tennessee 37996, USA}
\author{D.W.~Stracener}
\affiliation{\AddOakRidgePhys}
\author{M.~Mancuso}
\author{F.~Petricca}
\affiliation{Max-Planck-Institut f\"{u}r Physik, Munich D-80805, Germany}
\author{J.~Ninkovic}
\author{P.~Lechner}
\affiliation{MPG Semiconductor Laboratory, Munich D-80805, Germany}

\collaboration{KDK collaboration}\noaffiliation

\date{\today}

\begin{abstract}{
\Zn\ is a common calibration source, moreover used as a radioactive tracer in medical and biological studies. In many cases, $\gamma$-spectroscopy is a preferred method of \Zn\ standardization, which relies directly on the branching ratio of $J \pi (\Zn ) = 5/2^- \rightarrow J \pi (\Cu ) = 5/2^- $ via electron capture (\ECs ). We measure the relative intensity of this branch to that proceeding directly to the ground state (\EC ) using a novel coincidence technique, finding $\IEC / \IECs = \RhoResultFull $. Re-evaluating the decay scheme of \Zn\ by adopting the commonly evaluated branching ratio of $\Ibetap = 1.4271(7)\%$ we obtain $\IECs = \ECStarResultPercentFull $, and $\IEC = \ECResultPercentFull $. The associated \keV{1115} gamma intensity agrees with the previously reported NNDC value, and is now accessible with a factor of $\sim$2 increase in precision. Our re-evaluation removes reliance on the deduction of this gamma intensity from numerous measurements, some of which disagree and depend directly on total activity determination. The KDK experimental technique provides a new avenue for verification or updates to the decay scheme of \Zn , and is applicable to other isotopes.
}
\end{abstract}

\maketitle
 
\lhead{}
\chead{} 
\rhead{} 
\lfoot{}            
\rfoot{}                                                          
\renewcommand{\footrulewidth}{0.4pt}
\tableofcontents{}


\section{Introduction}

\Zn\ is a common $\gamma $-ray calibration source~\cite{hoppesNBSSpecialPublication1982}, and for nearly a century has been applied in the fields of medicine and biology as a radioactive tracer~\cite{banks1954estimation,hunt2002bioavailability,sugita2014evaluation}. It has been applied in various studies~\cite{lucconi2013use,korinko2013analysis} including an investigation of potential orbital-modulation effects on decay constants~\cite{pomme2016decay}.

In many applications, $\gamma$-ray spectroscopy is a convenient avenue for activity determination of \Zn , which is an emitter of essentially monoenergetic $\gamma$ rays (\keV{1115}) associated with some of its electron capture (EC) decays. This technique relies on knowledge of the absolute \keV{1115} intensity (fraction of decays that emit \keV{1115} $\gamma$s), available from existing decay scheme evaluations such as those by the \LNHB ~(LNHB)~\cite{TabRad_v3} or \NNDC ~(NNDC)~\cite{BROWNE20102425}. Though both evaluations report $\IkeV{1115}\sim 50\%$ with relative errors of $ \sim 0.2\%$, values (Table~\ref{tab:Intensity1115_LNHB_NNDC}) deviate by $\sim 0.4\%$ between the two sources. These evaluations combine dedicated measurements from the Euromet-721 exercise~\cite{Euromet2005, BE20061396} with other reported values (e.g. from Refs.~\cite{IWAHARA2005107,LUCA2003607}). All such determinations of absolute intensity are directly reliant on activity measurements, which may require various corrections; though the commonly used $4\pi\beta\gamma$ technique is influenced by low-energy X and Auger radiation when used with EC-decaying nuclides~\cite{funck1983influence}, such corrections have only been applied in some \Zn\ studies~\cite{BE20061396}. 

\begin{table}[ht]
    \centering
    \begin{tabular}{l u}
         &  \multicolumn{1}{c}{\IkeV{1115} (\%)} \\
        \hline\hline
     LNHB (2006)~\cite{TabRad_v3}    &  50.22 , 0.11  \\
     NNDC (2010)~\cite{BROWNE20102425}  &   50.04 , 0.10   \\
     \hline\hline
    \end{tabular}
    \caption{Absolute intensities of the \Zn\ \keV{1115} $\gamma$-ray, as reported in decay scheme evaluations of the \LNHB\ and \NNDC .}
    \label{tab:Intensity1115_LNHB_NNDC}
\end{table}

\begin{figure*}[ht]
    \centering
   \includegraphics[width=\textwidth]{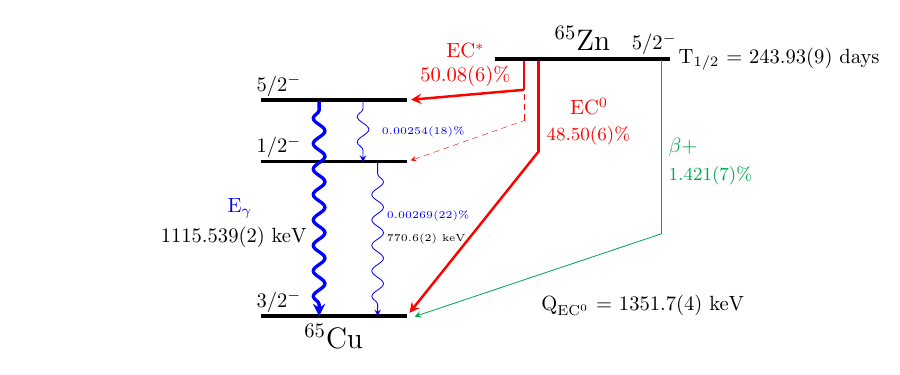}
    \caption{\label{fig:DecayScheme}The decay scheme of \Zn . Branching ratios are calculated combining our measurement of $\rho = \IEC / \IECs$ with the shown, commonly evaluated value of \Ibetap~\cite{TabRad_v3,BROWNE20102425} and assuming these three branches complete the decay scheme ($1=\IECs + \IEC + \Ibetap$). Gamma energies are from Ref.~\cite{HELMER200035} (\keV{1115.539(2)}), and adopted $\gamma $ levels~\cite{BROWNE20102425} (\keV{770.6(2)}). Intermediate $\gamma $ intensities are from Ref.~\cite{Euromet2005}. Total half-life is from Ref.~\cite{BROWNE20102425}, and $\text{Q}_{\text{EC}^0}$ is from \textsc{AME 2020}~\cite{AME2020_Wang_2021}.}
\end{figure*}

In this work, we present a novel measurement and resulting alternate, precise determination of \IkeV{1115} through re-evaluation of the decay scheme of \Zn. Using data from the KDK experiment~\cite{distefanoKDKPotassiumDecay2020,STUKEL2021165593}, \keV{1115}-producing electron capture decays ($ \ECs $) of \Zn\ are distinguished from those proceeding directly to the ground state ($ \EC $) of its Cu daughter. The KDK collaboration has recently employed this technique to obtain the first measurement of the exceedingly rare \EC\ decay of \K ~\cite{KDK_PRL, KDK_PRC}. The measurement of this work obtains a ratio of \Zn\ intensities, $\rho = \IEC / \IECs $. The decay scheme of \Zn , evaluated with our measurement, is displayed in Fig.~\ref{fig:DecayScheme}.

\section{Methods}

The KDK apparatus was designed for the discrimination between electron capture branches proceeding through an excited daughter state to those transitioning directly to the ground state. Below, we briefly outline the experimental setup, which has been fully characterized elsewhere~\cite{STUKEL2021165593}, with focus on features visible in the \Zn\ data and subsequently all physical processes relevant for obtaining a measurement of \IEC / \IECs .

\subsection{Apparatus and Visible Features}

\begin{figure}[ht]
    \centering
    \includegraphics[width=\linewidth]{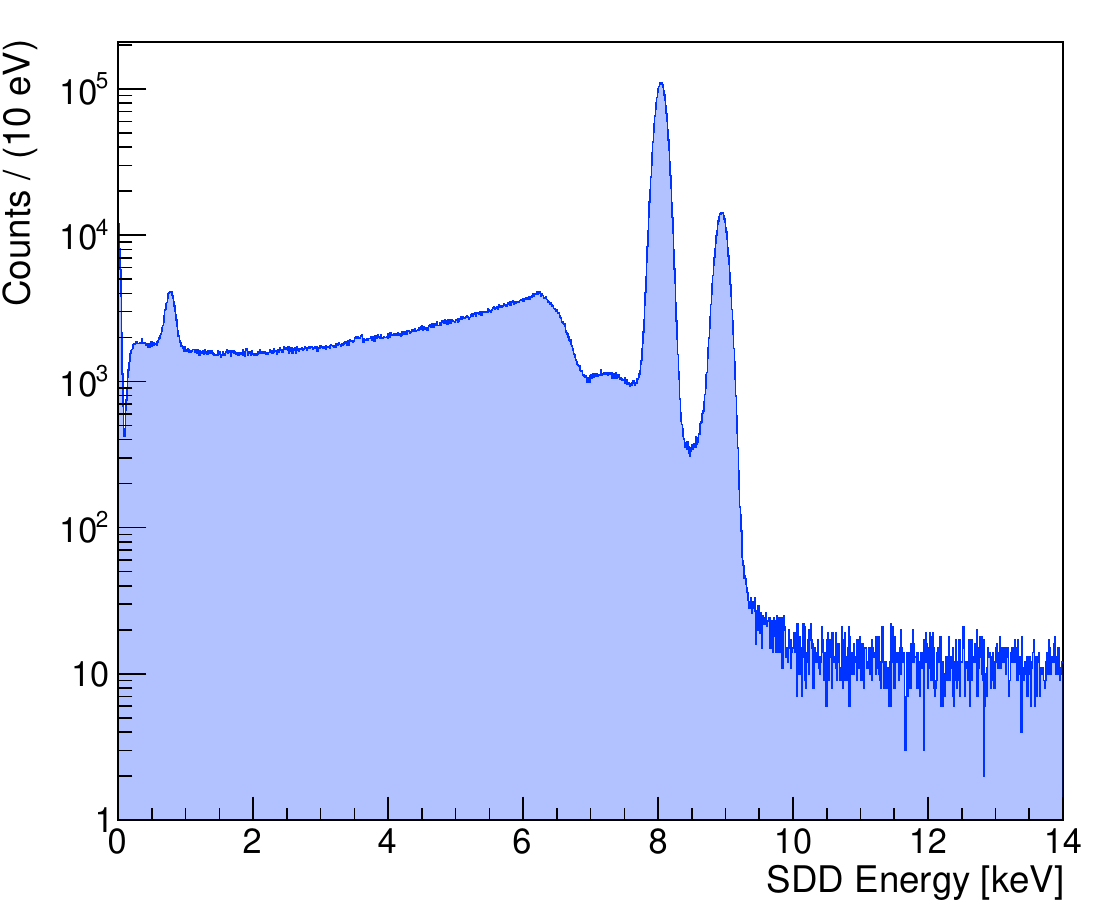}
    \caption{SDD spectrum obtained with the \Zn\ source over 1.4~days. The main features of interest are Cu \Kalpha\ and \Kbeta\ X-rays at 8 and 9~keV corresponding to \Zn\ electron captures. Some L-shell X-rays are visible below \keV{1}. A fit to SDD data distinguishes components near the K X-ray peaks in Fig.~\ref{fig:Zn_Fit}.}
    \label{fig:Zn_full_SddSpectrum}
\end{figure}

An inner Silicon Drift Detector (SDD) tags X-rays accompanying the source EC decays of interest, as shown in Fig.~\ref{fig:Zn_full_SddSpectrum}. Due to its excellent resolution (FWHM of \eV{200} at \keV{8}), the Cu \Kalpha\ (\keV{8}) and \Kbeta\ (\keV{8.9}) X-rays of interest are easily distinguishable. In the same region, the majority of remaining counts are attributed to K Augers from the same electron captures, and a flat background originating primarily from the source \betap\ branch. Below the signal (Cu) peaks, a small contribution from Zn \Kalpha\ X-ray fluorescence (\keV{8.6}) is included when fitting SDD spectra, as shown further. Moreover, due to the low noise threshold of the detector, L-shell X-rays (Cu, or fluoresced Zn) are visible near \keV{0.9}.

\begin{figure}[ht]
    \centering
    \includegraphics[width=\linewidth]{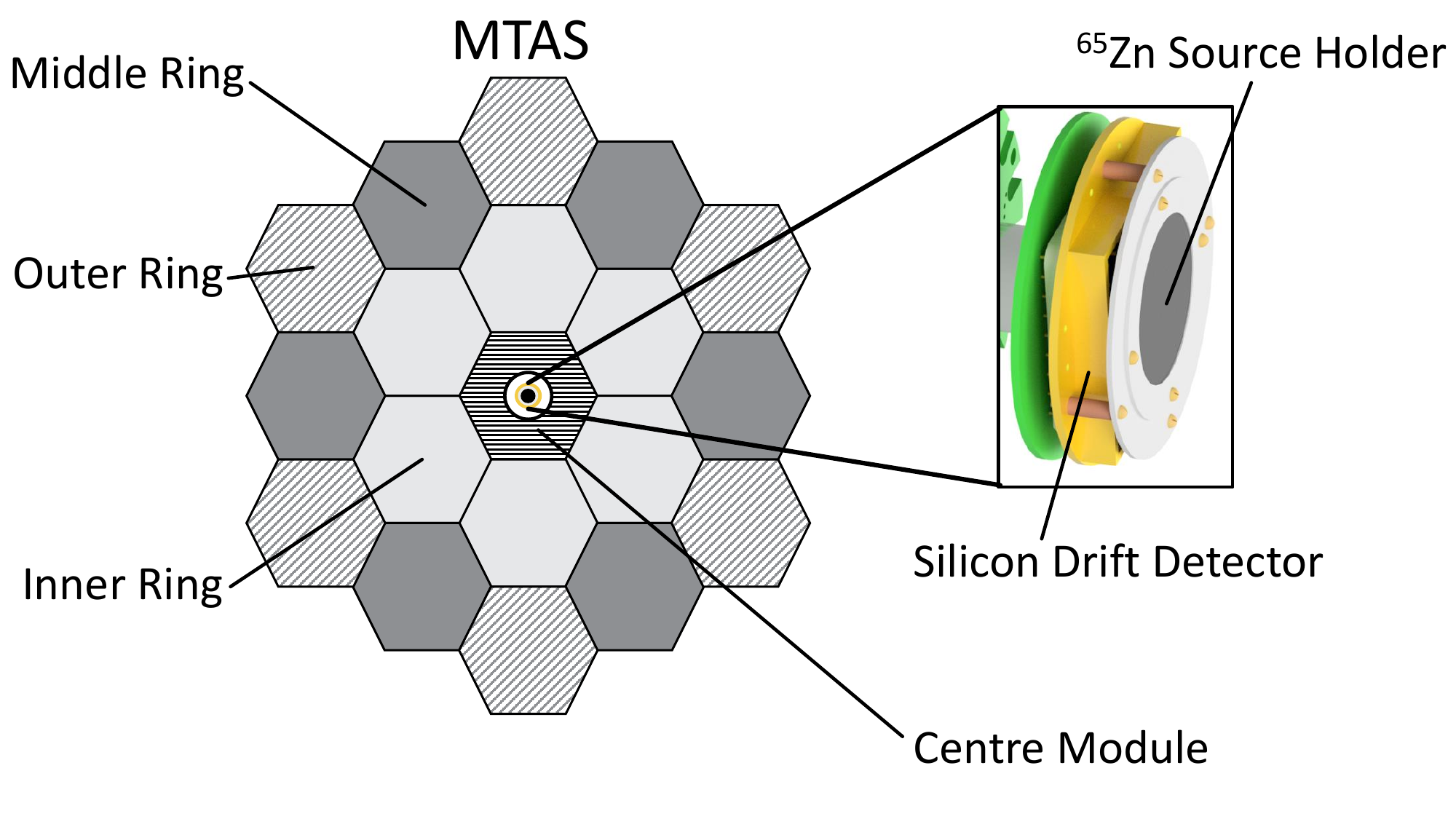}
    \caption{Schematic of the Silicon Drift Detector ($\mathcal{O}(\si{\square\milli\meter})$) and source placed inside the Modular Total Absorption Spectrometer ($\mathcal{O}(\si{\square\meter})$; cross-section displayed). The latter contains a central module along with inner, middle and outer rings, each containing NaI(Tl) volumes coupled to PMTs.}
    \label{fig:SddMtasSchematicSimple}
\end{figure}

The \Zn\ source and SDD are centered inside a large, outer Modular Total Absorption Spectrometer (MTAS) as depicted in Fig.~\ref{fig:SddMtasSchematicSimple}. MTAS is an extremely efficient NaI(Tl) $\gamma $-tagger, originally designed to study $\beta$-strength distributions of fission products~\cite{Rasco2015,KARNY201683}. For the \Zn\ $\gamma $ of interest (\keV{1115}), MTAS boasts a $\sim 98\%$ tagging efficiency~\cite{STUKEL2021165593}.

\begin{figure*}[ht]
    \centering
    \includegraphics[width=\linewidth]{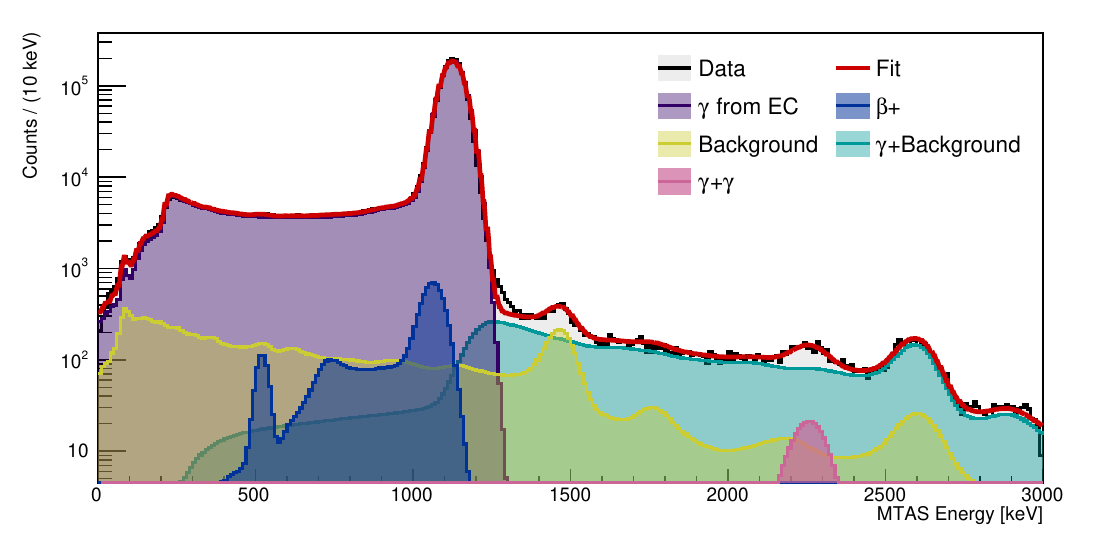}
    \caption{MTAS events of the \Zn\ run in a \mus{2} coincidence window with a SDD trigger. The dominant component is the \keV{1115} $\gamma $ spectrum, associated with the \Zn\ EC decays of interest. The measured MTAS background, and its convolution with the \keV{1115} spectrum are included. The \betap\ spectrum and a $\gamma + \gamma $ convolution provide additional contributions. Simulated $\gamma $ and \betap\ spectra are calibrated to data in energy and resolution.}
    \label{fig:MTAS_fit}
\end{figure*}

Several processes are visible in \Zn\ MTAS spectra, such as those in Fig.~\ref{fig:MTAS_fit}. Fully collected \keV{1115} $\gamma$s from \Zn\ \ECs\ decays form the most-prominent peak, though others originating from the natural MTAS background along with source+background and source+source sum-peaks are visible. The data is fit with calibrated, simulated spectra of \Zn\ $\gamma$s and \betap\ in MTAS, along with a measured MTAS background and convolutions. The shape of each spectrum is fixed, while the integral is allowed to vary. Such spectral analysis verifies simulation methods used to obtain MTAS efficiencies, and has been used to explore the source-SDD geometry~(Ref.~\cite{STUKEL2021165593} and App.~\ref{app:subsec:physical_quantities}).

SDD data, including the K X-ray electron-capture signal of interest, are categorized by coincidence with MTAS, generally using a $t^\prime = \mus{2}$ nominal coincidence window (CW). Details of coincidence characterization are available in our previous work~\cite{STUKEL2021165593}. The sorted SDD data is used to inform the parameter of interest, $\rho$, as described in the following section. 

\subsection{Physical Processes and Likelihood}\label{subsec:PhysicalProcesses_Likelihood}

The main result is obtained through a simultaneous minimization of the Baker-Cousins likelihood~\cite{BAKER1984437} on coincident and anti-coincident SDD spectra,
\begin{equation}
    -\ln \mathcal{L} = \sum_i \left\{ f_i( \boldsymbol{\theta}) - n_i + n_i\ln \left[ \frac{n_i}{f_i( \boldsymbol{\theta})} \right]  \right\},
    \label{eq:Likelihood_BakerCousins}
\end{equation}
which compares the total observed events ($n_i$) in a bin ($i$) to the corresponding model-predicted events ($f_i$). The parameters $\boldsymbol{\theta }$ include that of interest, $\rho$, along with various fixed and free terms pertaining to the spectra.

\begin{figure}[ht]
    \centering
    \includegraphics[width=\linewidth]{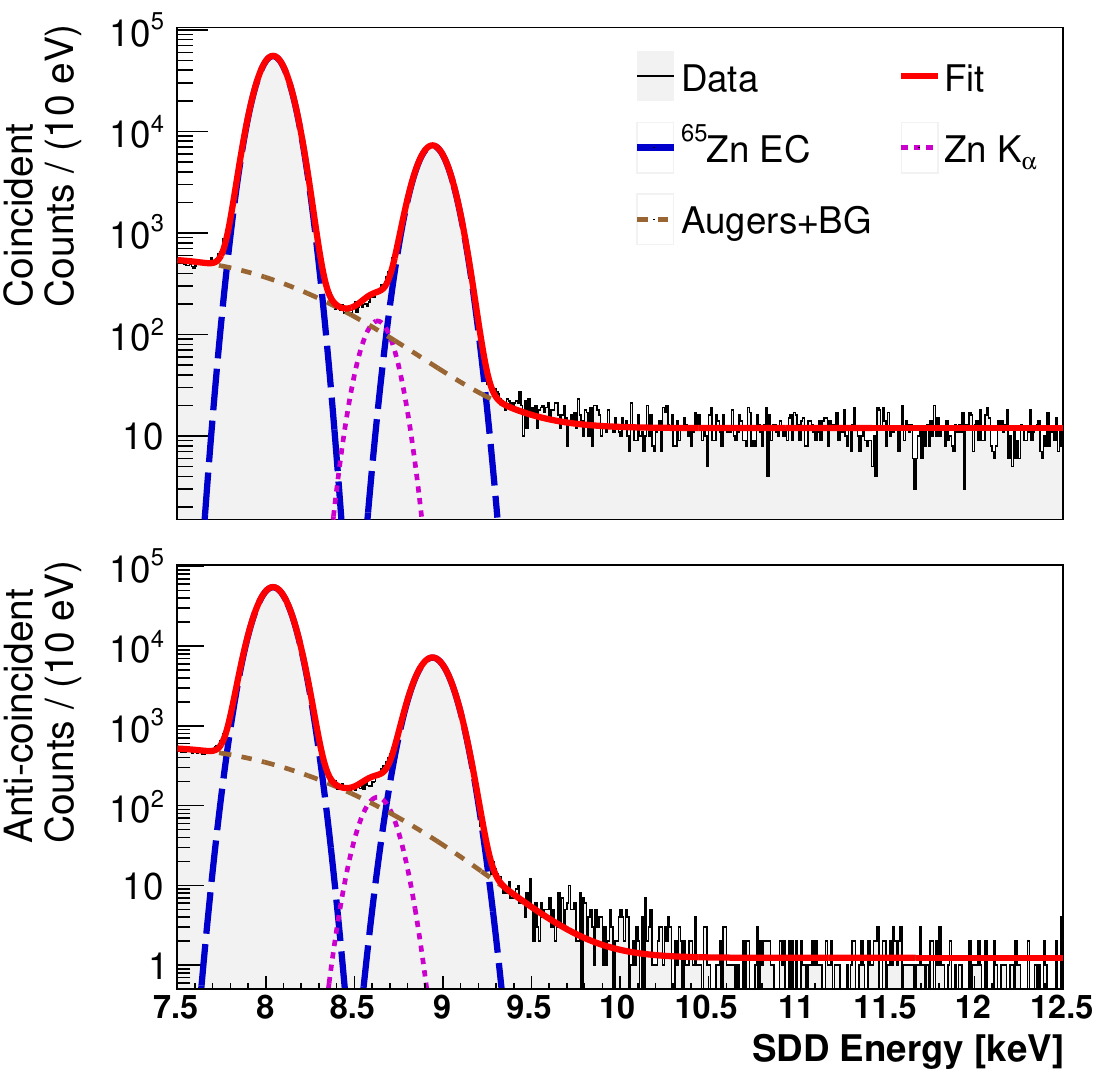}
    \caption{\label{fig:Zn_Fit}A fit to \Zn\ SDD data sorted using a \mus{2} coincidence window with MTAS. Cu \Kalpha\ and \Kbeta\ peaks from  \Zn\ electron captures inform $\rho = \IEC / \IECs $. Fluoresced Zn \Kalpha\ and a continuous component consisting of Augers and flat background complete the fit.}
\end{figure}

Such a simultaneous fit to data sorted using a \mus{2} CW is shown in Fig.~\ref{fig:Zn_Fit}. The main features are the Gaussian Cu \Kalpha\ and \Kbeta\ peaks of fixed means near 8 and \keV{9}. An additional X-ray peak near \keV{8.6} corresponds to fluoresced source Zn \Kalpha s. Lastly, an ad hoc component consists of a wide Gaussian term of free mean and width describing Cu K Augers, and a flat term attributed primarily to source \betap . The shape of all components is shared across coincident and anti-coincident spectra, whereas the integral of each can vary.

To obtain $\rho$, all processes affecting SDD signal detection and coincident categorization must be considered. To assess the former, we consider expected true, detected signal counts from the \ECs\ branch ($\sigma^*$) along with those from the \EC\ branch ($\sigma^0$),
\begin{align}
    \sigma^* & = \mathcal{N}\IECs P_K^* \omega_K (1 - \eta_\gamma )\eta 
    \nonumber
    \\
    \sigma^0 & = \mathcal{N}\IEC P_K^0 \omega_K \eta .
    \label{eq:sigmastar_sigma_expressions_main}
\end{align}
Both expressions above require the production of a Cu K X-ray and its successful detection in the SDD. To first order, the total source decays throughout data collection ($\mathcal{N}$), fluorescence probability ($\omega_K$), and SDD tagging efficiency of Cu K X-rays ($\eta$) do not need to be known to inform $\rho$,
\begin{align}
    \rho & = \frac{\IEC }{\IECs }
    \nonumber
    \\
    & = \frac{\sigma^0 }{\sigma^* }\frac{P_K^*}{P_K^0} (1 - \eta_\gamma ).
    \label{eq:rho_with_Kcap_etaG}
\end{align}
However, K-shell capture probabilities ($P_K$) differ between the two electron capture branches, and in the case of \ECs - originating signal, we require that the associated de-excitation gamma is not also seen in the SDD ($1 - \eta_\gamma$). K-capture probabilities are obtained from BetaShape~V2.2~\cite{mougeot2017betashape}, and $\eta_\gamma $ is obtained combining a measured geometric efficiency with simulations (App.~\ref{app:subsec:physical_quantities}). Above, \amend{internal conversion, internal pair formation, and internal Bremsstrahlung} have been neglected as these sub-dominant processes are negligible at our precision.

The expected true signal counts are related to the observed, sorted signal counts (those in the Cu K X-ray peaks of Fig.~\ref{fig:Zn_Fit}) by accounting for any process affecting coincidence sorting. One such effect stems from the imperfect MTAS $\gamma$-tagging efficiency $\epsilon $, which primarily affects sorting of \ECs -originating signal. This efficiency has been characterized for multiple coincidence windows by scaling measured \Mn\ efficiencies in energy through simulations, and correcting for deadtime~\cite{STUKEL2021165593}, yielding a value of $97.93(6)\%$ at the \mus{2} CW.

Moreover, any process detectable in MTAS can occur in spurious coincidence with the Cu SDD signal. The natural MTAS background rate is accounted for by considering a Poisson probability $P^0_\mathrm{BG} = \exp (-\beta \bar{t})$ of no natural MTAS background events within the effective coincidence window $\bar{t}$, along with an analagous probability of no MTAS-detected source \keV{1115} $\gamma $s ($P^0_\gamma$). Expected MTAS background coincidence rates have been obtained elsewhere~\cite{STUKEL2021165593}, and $P^0_\gamma$ is obtained in App.~\ref{App:LikelihoodDetails}. An additional source of coincidences with source \betap\ has been explored and deemed negligible. All spurious coincidences are proportional to the $\mathcal{O}(\si{\micro\second})$ CW, and, most significantly, place some \EC -originating signal in the coincident spectrum. Neglecting these terms thus tends to underestimate $\rho$, with increasingly dramatic effects at larger CWs.

With the above MTAS gamma efficiency and spurious coincidence considerations, expected coincident ($\Sigma^*$) and anti-coincident ($\Sigma^0$) signal counts are obtained,
\begin{align}
    \Sigma^* & = \frac{\nu}{1 + \rho^\prime } \biggl[ \epsilon + (1 - P^0_\mathrm{BG}P^0_\gamma)(1 - \epsilon + \rho^\prime ) \biggr]
    \nonumber
    \\
    \Sigma^0 & = \frac{\nu}{1 + \rho^\prime } P^0_\mathrm{BG}P^0_\gamma (1 - \epsilon + \rho^\prime ), 
    \label{eq:Expected_signal_counts_measured_maintext}
\end{align}
with more detail available in App.~\ref{App:LikelihoodDetails}. Above, $\nu = \sigma^* + \sigma^0$ are total signal events and $\rho^\prime = \sigma^0 / \sigma^*$ is introduced for simplicity ($ \rho^\prime \propto \rho$ via Eq.~\eqref{eq:rho_with_Kcap_etaG}). These expressions for expected signal counts are inserted directly into the model $f$ of Eq.~\eqref{eq:Likelihood_BakerCousins} as the integrals of Gaussian Cu K X-rays, such that $\rho$ is obtained directly from the fit along with its statistical error.

\section{Results}

\begin{figure}[ht]
    \centering
    \includegraphics[width=\linewidth]{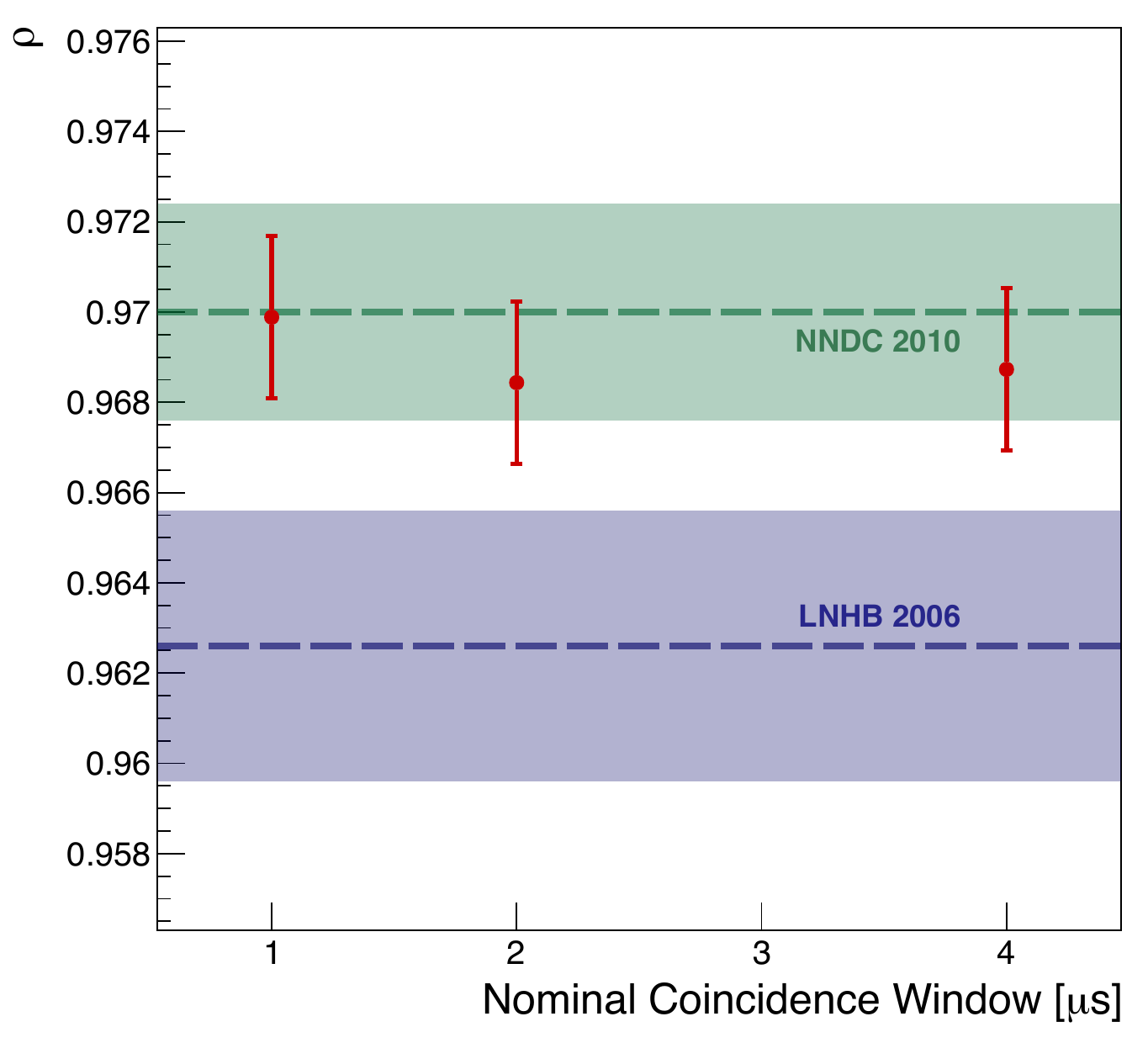}
    \caption{Measurements of $\rho = \IEC / \IECs$ obtained at three different coincidence windows (red points) are compared to expected values based on most-recent LNHB~\cite{TabRad_v3} and NNDC~\cite{BROWNE20102425} evaluations. The uncertainties on our measurements are correlated.}
    \label{fig:Zn_Rho_vs_CW}
\end{figure}

With the likelihood method described above, we obtain
\begin{equation}
    \rho = \IEC / \IECs = \RhoResultSepErrs ,
    \label{eq:RhoResult}
\end{equation}
using the \mus{2} dataset. This result, and those obtained using 1 and \mus{4} CWs are shown in Fig.~\ref{fig:Zn_Rho_vs_CW}. Our values of \IEC /\IECs\ agree across coincidence windows, and we note that these measurement uncertainties are correlated.
We also report values derived from branching ratios in existing NNDC~\cite{BROWNE20102425} and  LNHB~\cite{TabRad_v3} evaluations.

The systematic errors considered in this analysis generally fall into two categories: (1) physical limitations (finitely known quantities such as tagging efficiency; App.~\ref{app:subsec:physical_quantities}) and (2) spectral characteristics (e.g. fit range). The former (1) are accounted for analytically using equations of Sec.~\ref{subsec:PhysicalProcesses_Likelihood}. All physical parameters are varied simultaneously to account for covariances, and are assumed to follow Gaussian distributions. The latter (2) are gauged by performing numerous fits, such as that in Fig.~\ref{fig:Zn_Fit}, while randomly varying the binning and fit range. The final systematic error of $1.3\times 10^{-3}$ sums those of category (1) and (2) in quadrature, and within rounding happens to be equivalent to the statistical error. We find that the physical systematics completely dominate any spectral effects, with the leading contribution stemming from the error on MTAS $\gamma$-tagging efficiency. Details of these considerations are provided in App.~\ref{app:SystErrors}, and contributions of individual sources of systematic error are summarized in Table~\ref{tab:SystErrors}.

We re-evaluate the decay scheme of \Zn\ (Fig.~\ref{fig:DecayScheme}) by combining our result of \IEC / \IECs\ with the commonly evaluated branching ratio $\Ibetap = 1.421(7)\%$~\cite{BROWNE20102425,TabRad_v3}, and assuming these three branches complete the decay scheme. \amend{Our result is not sensitive to internal conversion and pair formation, internal Bremsstrahlung}, de-excitation through, or EC to the intermediate \keV{770} Cu level. The evaluated \IECs\ is thus equivalent to the absolute \keV{1115} $\gamma $ intensity. 

\begin{figure}[ht]
    \centering
    \includegraphics[width=\linewidth]{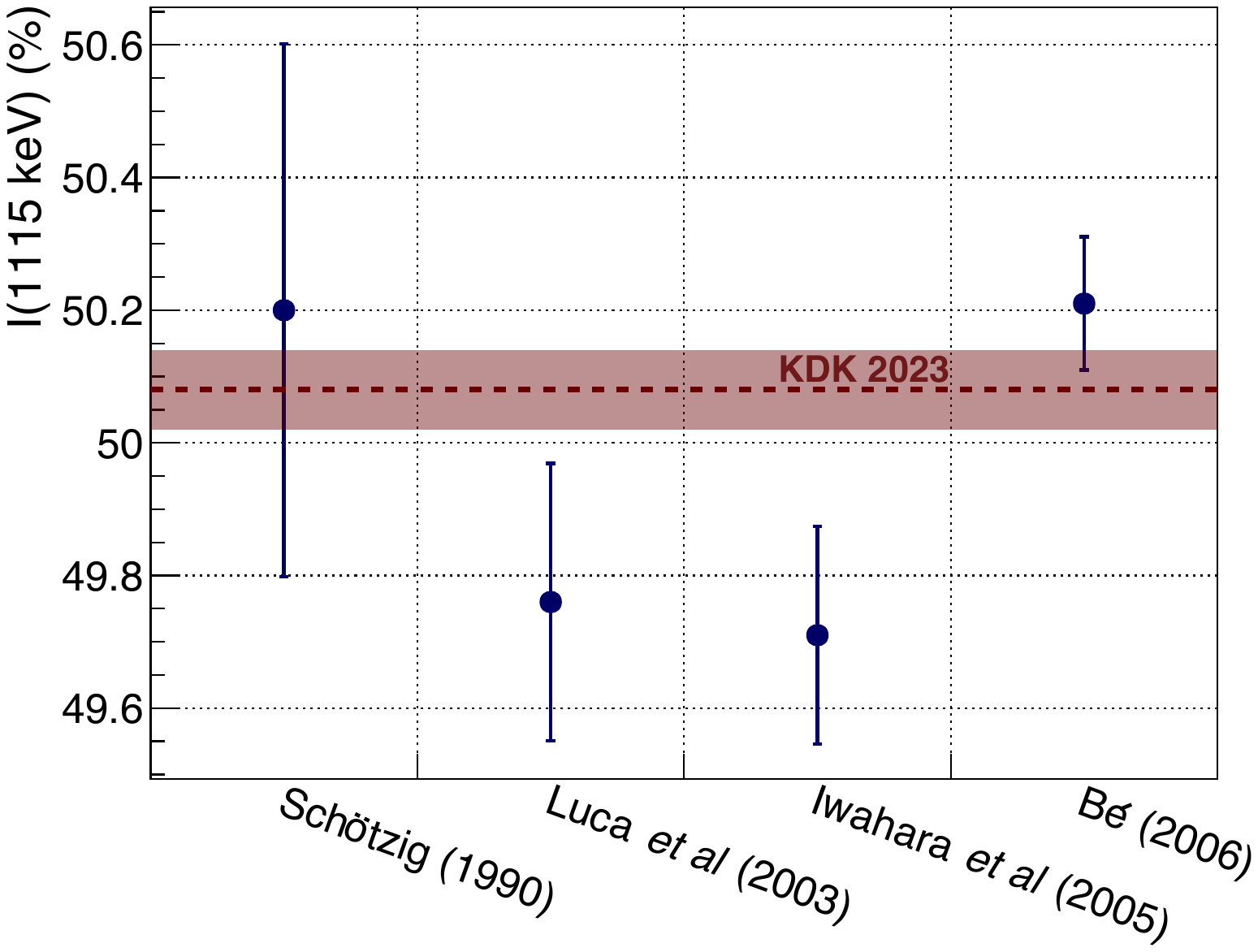}
    \caption{Selected measured absolute intensities of the \Zn\ \keV{1115} gamma-ray (blue points) along with the re-evaluation of this work (dark red line).
    Individual measurements~\cite{SCHOTZIG1990523,LUCA2003607,IWAHARA2005107} and a collective result from Euromet~\cite{BE20061396,Euromet2005} are shown, with values available in Table~\ref{tab:GammaIntensityValues}. Additional measurements are listed in the evaluations of Refs.~\cite{TabRad_v3,BROWNE20102425}.}
    \label{fig:IECs_measurements_and_KDK}
\end{figure}

With our re-evaluation, we find an improvement in sensitivity to the \keV{1115} branching ratio of almost a factor of 2 relative to that obtained by the Euromet exercise~\cite{Euromet2005,BE20061396}. We compare this result of $\IkeV{1115} = 50.08(6)\%$ to various most-recent measurements in Fig.~\ref{fig:IECs_measurements_and_KDK}, with values listed in Table~\ref{tab:GammaIntensityValues}. An alternate re-evaluation combining the result of this work with a theoretical $\IEC / \Ibetap $ ratio from Betashape~V2.2~\cite{mougeot2017betashape} yields highly consistent results (Table~\ref{tab:DecayScheme_ReEval}).

\section{Conclusions}
\label{Sec:Conclusion}

We have re-determined the absolute emission intensity of \keV{1115} $\gamma$s originating from \Zn\ electron-capture decays, improving the available precision by a factor of $\sim $2, with a result that agrees with the existing National Nuclear Data Center evaluation. This improvement stems from precise determination of a ratio of electron-capture decay intensities, which does not require precise knowledge of source activity. As in other experiments, the efficiency of our $\gamma$-tagger is a limiting systematic, though this precisely known quantity yields a systematic error roughly equivalent to our statistical uncertainty. Additionally, our determination is the first involving the total \EC\ branching ratio of \Zn . Our successful measurement is analogous to an existing result for \K , and may be applied to other nuclides which decay through multiple modes of electron capture. The unique detector configuration of the KDK experiment provides a precise result available for use in the decay scheme evaluation of the familiar and vastly used \Zn .


\acknowledgments

We thank Xavier Mougeot and Sylvain Leblond of LNHB for input on  \Zn. Work was performed at Oak Ridge National Laboratory, managed by UT-Battelle, LLC, for the U.S. Department of Energy under Contract DE- AC05-00OR22725. The United States Government retains and the publisher, by accepting the article for publication, acknowledges that the United States Government retains a non-exclusive, paid-up, irrevocable, world-wide license to publish or reproduce the published form of this manuscript, or allow others to do so, for United States Government purposes. The Department of Energy will provide public access to these results of federally sponsored research in accordance with the DOE Public Access Plan (\href{http://energy.gov/downloads/doe- public-access-plan}{http://energy.gov/downloads/doe- public-access-plan}). US support has also been supplied by the Joint Institute for Nuclear Physics and Applications, and by NSF grant EAR-2102788. Funding in Canada has been provided by NSERC through a SAPIN  grant, as well as by the Faculty of Arts and Science of Queen's University, and by the McDonald Institute.

\clearpage
\bibliography{References}

\clearpage
\appendix

\section{Components of the \texorpdfstring{\Zn}{65Zn} Decay Scheme and Re-evaluated Branching Ratios}\label{App:ReEval_DecayScheme}

The decay scheme of \Zn\ (Fig.~\ref{fig:DecayScheme}) can be constructed either wholly empirically, or with a combination of measurements and theoretical values. Prior to this work, the purely empirical evaluation relied on intensity measurements of \keV{1115} and \keV{511} $\gamma $s associated with \Zn .

The \keV{1115} $\gamma $ intensity is equivalent to the branching ratio to the \keV{1115} Cu state (\IECs ) assuming a simplified scheme where (1) the intermediate 344, 770~keV $\gamma$ emissions are negligible, and (2) internal conversion is negligible. The order of precision for \IECs\ from both previous evaluations~\cite{TabRad_v3,BROWNE20102425} is $\sim \mathcal{O}(10^{-3})$, and the $\mathcal{O}(10^{-5})$ intensities of intermediate $\gamma$s~\cite{Euromet2005} along with the $\mathcal{O}(10^{-4})$ internal conversion process (\textsc{BrIcc} program~\cite{KIBEDI2008202}) can be neglected. \amend{Radiative electron capture (REC) may accompany both the \EC\ and \ECs\ \Zn\ decays, resulting in internal Bremsstrahlung emission. However, for these allowed transitions, the frequency of REC relative to non-REC decays is of the order of $ 10^{-4}$~\cite{bambynek1977orbital} and can be neglected in the determination of \IECs .}

The branching ratio through \betap\ disintegration (\Ibetap ) is informed from measured \keV{511} intensities obtained taking annihilation-in-flight~\cite{colgate1953electron} into account. The decay scheme is then built with
\begin{equation}
    1 = \IECs + \IEC + \Ibetap ,
\end{equation}
which yields the final required branching ratio (\IEC ). To re-evaluate the scheme with the result of this work, the above assumption of unitarity is maintained and the measured parameter $\rho = \IEC / \IECs$ is inserted,
\begin{equation}
    1 = \IECs (1 + \rho ) + \Ibetap .
\end{equation}
Above, $\Ibetap = 1.4271(7)\%$~\cite{TabRad_v3,BROWNE20102425} can be fixed to obtain \IECs\ and subsequently \IEC . We assume this \Ibetap\ value is effectively uncorrelated with the EC branching ratios. Alternatively, $\rho$ may be combined with a theoretical value of $\rho_B = \IEC  / \Ibetap = 33.4(7)$ from Betashape V2.2~\cite{mougeot2017betashape},
\begin{equation}
    1 = \IECs \left( 1 + \rho + \rho \frac{1}{\rho_B} \right),
\end{equation}
which yields consistent branching ratios. These re-evaluations are compared to existing ones in Table~\ref{tab:DecayScheme_ReEval}. Both the main and alternate methods above reduce the uncertainty on \Zn\ EC branching ratios by a factor of 2 compared to existing evaluations~\cite{TabRad_v3,BROWNE20102425}. The re-evaluated intensity of the \keV{1115} $\gamma$ (=\IECs ) is compared to other measurements in Table~\ref{tab:GammaIntensityValues} and Fig.~\ref{fig:IECs_measurements_and_KDK}. Calculations with our measurement remain insensitive to internal conversion and intermediate gamma emissions. \amend{Radiative electron capture does not affect our measurement of $10^{-3}$ precision. Internal Bremsstrahlung photons are detectable in both the SDD and MTAS, and can create false positives and negatives. All such effects are suppressed by $\mathcal{O}( 10^{-4})$ relative to each radiative electron capture branch. In the next generation of precision measurements, internal Bremsstrahlung and other low-order effects will contribute.}

\begin{table*}[ht]
    \centering
    \begin{tabular}{l....}
            &   \multicolumn{1}{c}{LNHB~(2006)} &   \multicolumn{1}{c}{NNDC~(2010)} &   \multicolumn{1}{c}{This work (main)}    &   \multicolumn{1}{c}{ This work (alt.) }   \\
        \hline\hline
        \IECs\ (\%)    &  50.23(11) &   50.04(10)   &   \ECStarResultPercentFullConciseALT  &     50.06(5)  \\
        \IEC\ (\%)    &   48.35(11) &   48.54(7)    &   \ECResultPercentFullConciseALT  &     48.48(5)    \\
        \Ibetap\ (\%)   &   1.421(7)   &   1.421(7) &   1.421(7)    &    1.45(3)    \\
        \hline\hline
    \end{tabular}
    \caption{Branching ratios of \Zn\ decays from existing evaluations reported by the LNHB~\cite{TabRad_v3} and NNDC~\cite{BROWNE20102425} along with the re-evaluation of this work. The main re-evaluation combines the listed, adopted value of \Ibetap\ with the KDK measurement of $\IEC / \IECs = \RhoResultFull $. The alternate re-evaluation combines the KDK measurement with the theoretical ratio of $\IEC / \Ibetap = 33.4(7)$ from Betashape V2.2~\cite{mougeot2017betashape}, leading to a larger error on the resulting \Ibetap\ value relative to the $\sim 4$ times more-precise, measured \Ibetap .}
    \label{tab:DecayScheme_ReEval}
\end{table*}

\begin{table}[ht]
    \centering
    \begin{tabular}{ l . . }
        Source  &   \multicolumn{1}{c}{$I(\keV{1115})$ (\%)} &   \multicolumn{1}{c}{Rel. error (\%)}  \\
    \hline\hline
        Sch\"{o}tzig~(1990)~\cite{SCHOTZIG1990523}  &   50.2    &   0.8    \\
        Luca~\etal~(2003)~\cite{LUCA2003607}    &  49.76   &  0.42     \\
        Iwahara~\etal~(2005)~\cite{IWAHARA2005107}    &   49.71  &   0.33    \\
        B\'{e}~(2006)~\cite{BE20061396,Euromet2005} &   50.21   &   0.20    \\
        This work   &  \ECStarResultPercent   &   \ECStarResultPercentRelError       \\
    \hline\hline
    \end{tabular}
    \caption{Absolute intensities of the \keV{1115} $\gamma$-ray of \Zn\ from existing measurements along with the value deduced from the decay scheme re-evaluation of this work (Fig.~\ref{fig:DecayScheme} and App.~\ref{App:ReEval_DecayScheme}).}
    \label{tab:GammaIntensityValues}
\end{table}

\section{Likelihood details}\label{App:LikelihoodDetails}

The likelihood fit performed in Fig.~\ref{fig:Zn_Fit} consists of 4 distinct spectral components in both the coincident and anti-coincident spectra. Counts in the Gaussian Cu \Kalpha\ and \Kbeta\ peaks corresponding to \Zn\ electron captures are used to inform $\rho$ as discussed in the main text, with some details available below. Symbol definitions are retained from the main text, where applicable.

\subsection{Coincidence sorting}

Expected coincident signal counts consist of the terms,
\begin{widetext}
\begin{equation}
    \Sigma^* =
    \overbrace{
        \sigma^* \epsilon 
    }^\text{\ECs , $\gamma$ detected}
    +
    \overbrace{
        \sigma^* (1 - \epsilon ) (1 - P^0_\mathrm{BG} P^0_\gamma ) 
    }^\text{\ECs , $\gamma $ missed , BG coincidence}
    +
    \overbrace{
        \sigma (1 - P^0_\mathrm{BG} P^0_\gamma )
    }^\text{\EC , BG coincidence}
    ,
\end{equation}
\end{widetext}
which contain \ECs -originating events whose gamma-ray was successfully tagged by MTAS, those where the gamma was missed occurring in coincidence with another event in MTAS, and lastly \EC -originating events which occurred with an event in MTAS. With the substitution $\sigma^* = \nu / (1 - \rho^\prime )$, this expression yields Eq.~\eqref{eq:Expected_signal_counts_measured_maintext} in the main text. The expression for expected anti-coincident signal counts $\Sigma^0$ is the complement to the above such that all events are accounted for: $\Sigma^* + \Sigma = \sigma^* + \sigma^0 = \nu$.

The primary source of spurious coincidences is the natural MTAS background of rate $\beta$, corrected for via
\begin{equation}
    P^0_\mathrm{BG} = e^{-\beta \bar{t}},
    \label{eq:P_0_BG}
\end{equation}
which is the probability of no occurrences within an average coincidence window $\bar{t}$. Additionally, there is a rate of \ECs\ events which are not detected in the SDD, though the associated $\gamma$-ray is detected in MTAS in coincidence with SDD signal. Such events have an expected rate
\begin{equation}
    \mathcal{R}_* = A \IECs (1 - \eta_* ) \epsilon ,
    \label{eq:R_ECStar}
\end{equation}
where $A$ is the source activity and a lack of SDD X-ray detection is ensured via
\begin{equation}
    (1 - \eta_* ) = P_K^* (1 - \omega_K \eta ) + (1 - P_K^* ).
    \label{eq:R_ECStarTerm_EtaStar}
\end{equation}
At the order of this correction, the $<1\%$ probability of $\gamma $ interaction with the SDD (App.~\ref{app:subsec:physical_quantities}) is negligible. This additional source of spurious coincidences has a 0-event probability within $\bar{t}$ of $P^0_\gamma = \exp (-R_* \bar{t})$. Altogether, the probability of any (1+) event(s) from $\beta$ or $\mathcal{R}_*$ within the timescale $\bar{t}$ is $(1 - P^0_\mathrm{BG} P^0_\gamma )$.

 Expected MTAS background counts $\beta \bar{t}$ have been measured directly~\cite{STUKEL2021165593}, and are far larger than those from $\mathcal{R}_*$, as shown in Table~\ref{tab:exp_spurious_coinc_counts}. To obtain $\mathcal{R}_*$, known values for K-capture~\cite{mougeot2017betashape} and fluorescence~\cite{schonfeld1996evaluation} probabilities are used, along with an assumed partial \ECs\ activity $A\IECs $ from calculated source activity~\cite{STUKEL2021165593} and $\IECs \sim 0.5$ with an assigned error of 10\%. MTAS $\gamma$ and SDD X-ray tagging-efficiencies have been measured elsewhere~\cite{STUKEL2021165593}, wherein average coincidence windows $\bar{t}$ are obtained from reported quantities $\beta\bar{t}, \beta$.

 \begin{table}[ht]
     \centering
     \begin{tabular}{l . .}
      CW ($\si{\micro\second}$)    &   \multicolumn{1}{c}{$\beta \bar{t}$ ($10^{-2}$)} &   \multicolumn{1}{c}{ $\mathcal{R}_* \bar{t}$ $(10^{-3}$) }  \\
      \hline\hline
      1 &   0.74(1)    &  0.7(1) \\
      2 &   1.25(1)    &   1.1(2)    \\
      4    &  2.27(2)   &  2.1(3) \\
          \hline\hline
     \end{tabular}
     \caption{Expected counts leading to spurious coincidences of SDD signal with MTAS events for a coincidence window $\bar{t}$. The natural MTAS background ($\beta\bar{t}$) dominates the effect of $\gamma$s from \ECs\ events which were missed by the SDD ($\mathcal{R}_* \bar{t}$).}
     \label{tab:exp_spurious_coinc_counts}
 \end{table}

\begin{figure}[ht]
    \centering
    \includegraphics[width=\linewidth]{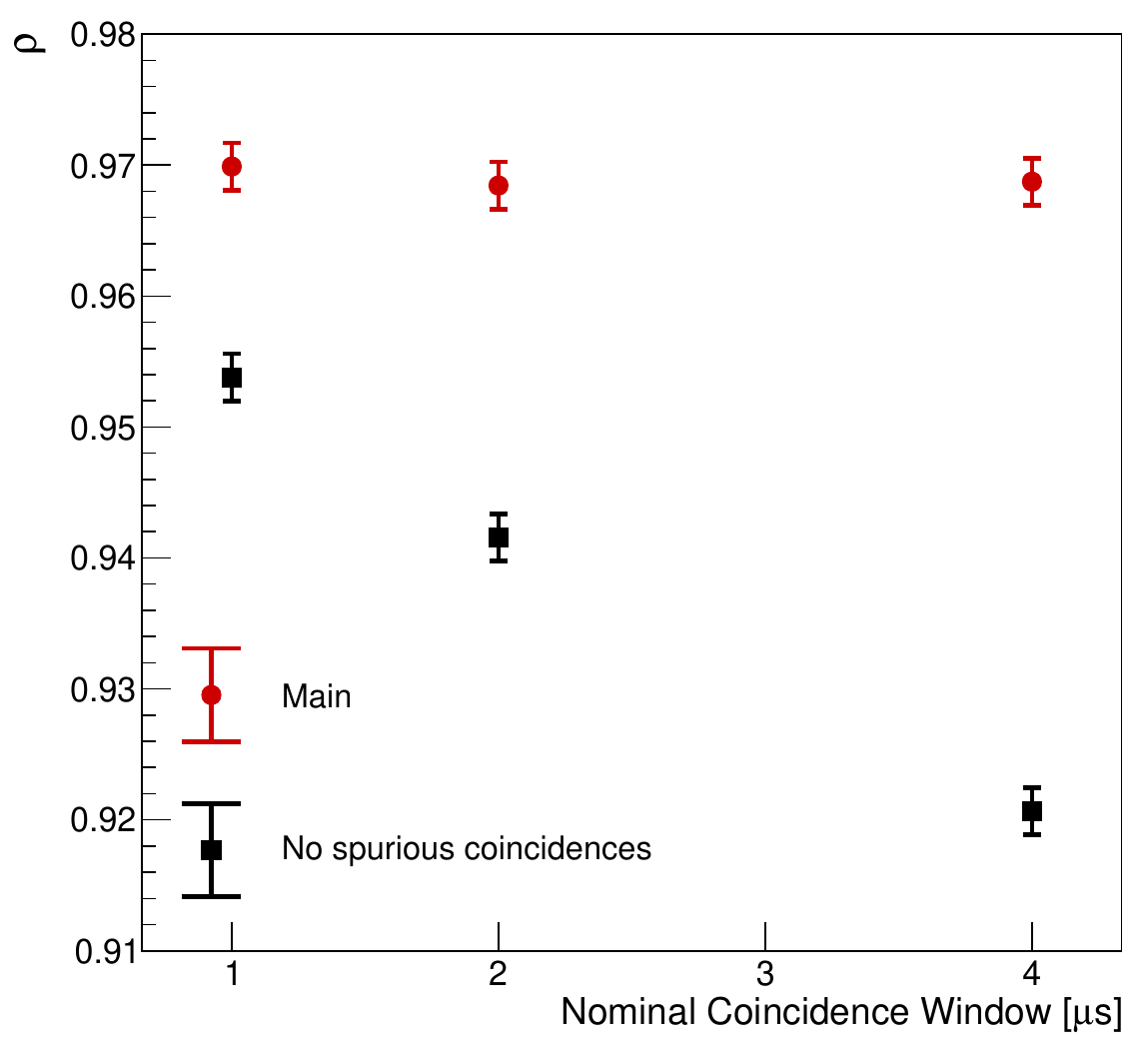}
    \caption{Obtaining $\rho =\IEC / \IECs $ of \Zn\ while neglecting expected spurious coincidences of signal with background (black squares) produces an unphysical, linear dependence on coincidence window. The corrected results (red circles) are consistent across CWs.}
    \label{fig:Rho_vs_cw_main_no_false_neg}
\end{figure}

 The effect of neglecting spurious coincidences is depicted in Fig.~\ref{fig:Rho_vs_cw_main_no_false_neg}. Without this correction, results for $\rho$ are directly anti-correlated with coincidence window. Applying the $\mathcal{O}(\text{CW})$ corrections for such coincidences resolves the unphysical behaviour. 

\subsection{Physical quantities}\label{app:subsec:physical_quantities}

Beyond coincidence sorting, any processes affecting relative production or SDD detection of \EC\ and \ECs\ events must be accounted for. As per Eq.~\eqref{eq:rho_with_Kcap_etaG}, $\rho $ is directly dependent on relative K-shell capture probabilities $P_K^* , P_K^0$, and SDD efficiency of tagging gammas $\eta_\gamma$. 

K-capture probabilities from Betashape V2.2~\cite{mougeot2017betashape} yield $P_K^0 / P_K^* = 1.00690(44) $. Due to the high precision of the measurement, this seemingly small variation in relative probabilities is the second-most-dominant source of systematic error, as discussed further. It is notable that previously reported~\cite{TabRad_v3} values from a 1996 evaluation~\cite{schonfeld1996evaluation} yield a consistent ratio of $P_K^0 / P_K^* = 1.0067(27)$.

The SDD $\gamma$-tagging efficiency is obtained from a mixture of measurement and simulations. The SDD K X-ray tagging efficiency obtained elsewhere~\cite{STUKEL2021165593} is assumed to be equivalent to the geometric efficiency of the SDD; 8-9~keV Cu X-rays penetrate the dead layers of the SDD without escaping the detector, and all are likely to escape the source, as same-energy Auger electrons are readily visible in the SDD (Fig.~\ref{fig:Zn_Fit}). Two terms are considered,
\begin{equation}
    \eta_\gamma = \eta \eta_\gamma^i,
\end{equation}
where $\eta = 21.5(11)\%$ is the geometric SDD efficiency~\cite{STUKEL2021165593}, and $\eta_\gamma^i$ is the probability that the $\gamma$ interacts with (does not escape) the SDD volume. 

Simulations are not relied upon for total SDD tagging efficiencies due to the uncertainty in sub-mm, geometric source-SDD modelling. As such, only the probability of $\gamma$ interaction with the volume it passes through is obtained through a simulation of 10~million events, yielding $\eta_\gamma^i = 1.522(6)\%$. The SDD $\gamma$-tagging efficiency is then $\eta_\gamma = 0.327(17)\%$.

\section{Systematic errors}\label{app:SystErrors}

Systematic errors are accounted for in two separate groups: physical sources (1) discussed in App.~\ref{app:subsec:physical_quantities}, and spectral characteristics (2) chosen prior to the likelihood fit.

Every physical source of error is included in expressions for expected coincident and anti-coincident signal counts of Eq.~\eqref{eq:Expected_signal_counts_measured_maintext}. Combined with Eq.~\eqref{eq:rho_with_Kcap_etaG}, $\rho^\prime = \sigma^0 / \sigma^*$ and spurious coincidence expressions (Eqs.~\eqref{eq:P_0_BG}--\eqref{eq:R_ECStarTerm_EtaStar}), $\rho$ is related to all physical parameters. Values of all such parameters are listed in Table~\ref{tab:Phys_syst_errors_parameters}.

\begin{table}[ht]
    \centering
    \begin{tabular}{ldl}
    Parameter   &   \multicolumn{1}{c}{Value}   &   Source  \\
    \hline\hline
     $P_K^*$ (\%)    &   87.497(28) &   Betashape V2.2~\cite{mougeot2017betashape}  \\
     $P_K^0$ (\%)    & 88.101(26)   &   Betashape V2.2~\cite{mougeot2017betashape}  \\
     $\omega_K$ (\%) &   45.4(4)    &   Ref.~\cite{schonfeld1996evaluation} \\
     $\eta$ (\%) &   21.5(11)   &   Ref.~\cite{STUKEL2021165593}    \\
     $\epsilon$ (\%) &   97.93(6)    &   Ref.~\cite{STUKEL2021165593}$^\dagger$ \\
     $\beta \bar{t}$    &   0.0125(1)    &   Ref.~\cite{STUKEL2021165593}$^\dagger$ \\
     $\beta$ (kHz)    &   2.63951(15)    &   Ref.~\cite{STUKEL2021165593}$^\dagger$ \\
     $A\IECs  $ (kBq)  &    0.268(29)   &   App.~\ref{app:subsec:physical_quantities} \\
     $\eta_\gamma^i$ (\%) &   1.522(6)  &   App.~\ref{app:subsec:physical_quantities}   \\
     \hline\hline
    \end{tabular}
    \caption{Values and errors of physical parameters which contribute to the overall systematic error. $^\dagger$Value for \mus{2} coincidence window, with others available within Ref.~\cite{STUKEL2021165593}.}
    \label{tab:Phys_syst_errors_parameters}
\end{table}

All physical parameters, such as the MTAS $\gamma$-tagging efficiency, are assumed to follow Gaussian distributions with a width corresponding to their 68\% CL error. The effect of this efficiency on $\rho$ is gauged by sampling 1,000 values from the efficiency distribution, and obtaining the resulting difference in obtained $\rho$ values. The width of the latter distribution corresponds to the systematic error due to MTAS $\gamma$-tagging efficiency, which is found to be $1.2\times 10^{-3}$. 

\begin{figure}[ht]
    \centering
    \includegraphics[width=\linewidth]{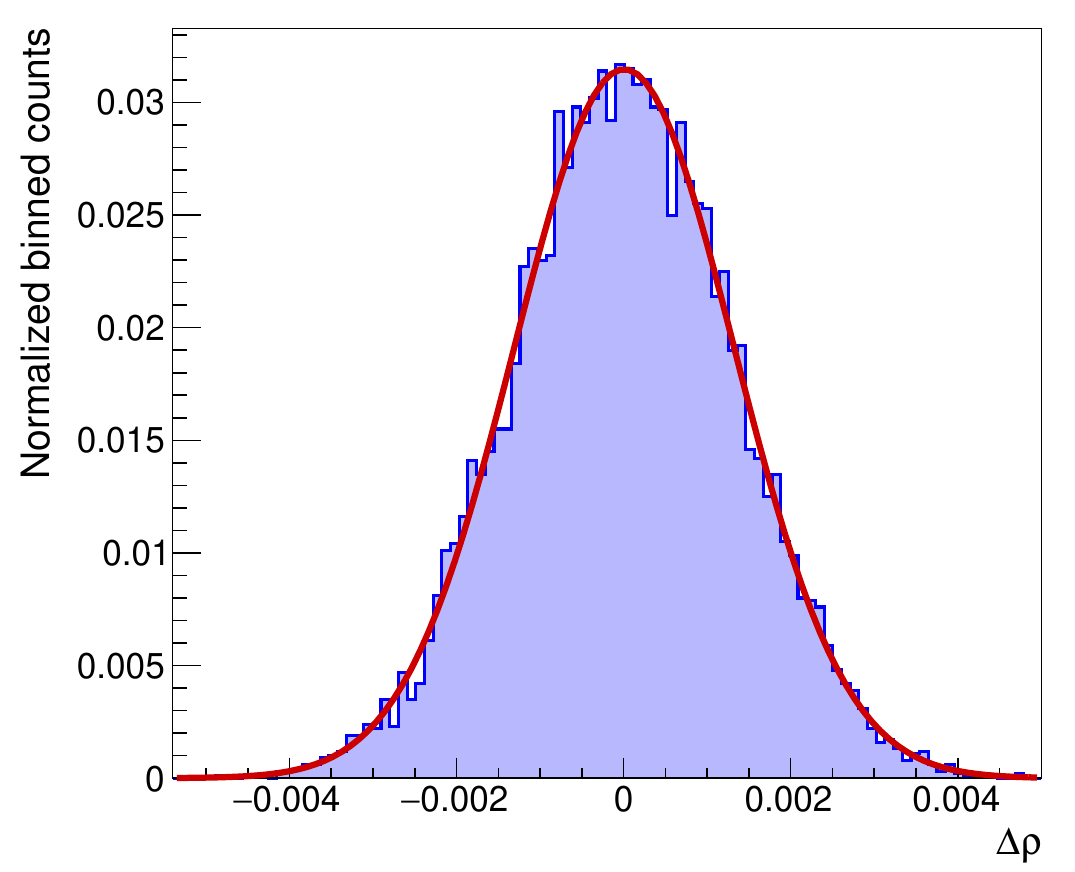}
    \caption{Variation in $\rho$ induced from \amend{10,000} instances of randomly varying all physical parameters used to obtain the result. The standard deviation of this distribution ($1.3 \times 10^{-3}$) corresponds to the systematic error on $\rho$. Binned counts are normalized to total iterations.}
    \label{fig:PhysicsSystematicsSpread}
\end{figure}

To account for covariances, all physical parameters are varied simultaneously over 10,000 iterations, resulting in a distribution of $\rho $ variation as shown in Fig.~\ref{fig:PhysicsSystematicsSpread}. The spread of this distribution, $1.33 \times 10^{-3}$, is equivalent to the total systematic error of category (1).

Remaining sources of systematics are of category (2), and contain the choice of binning and fit range. Bin widths were considered between $[10, 30]$~eV, ensuring X-ray peaks remain easily distinguishable. The low end of the fit range is constrained by the validity of the ad hoc model, as the wide Gaussian describing primarily Auger counts (of three different energy ranges) is not sufficient to describe data at much lower energies than pictured in Fig.~\ref{fig:Zn_Fit} (data at lower energies is shown in Fig.~\ref{fig:Zn_full_SddSpectrum}). Both the low and high energy cuts must encompass the ad hoc background in a region not dominated by X-rays. The sets of low and high cuts considered are $ [7.2, 7.5] \text{ keV} \times [10.5, 14.0] \text{ keV}$.  

The effect of the binning is gauged by performing fits while randomly varying the bin width from a uniform distribution bounded as described above. The resulting distribution of $\rho$ values has a width of $\mathcal{O}(10^{-5})$. The effect of varying the fit range is similarly small. The overall systematic error of category (2) is equivalent to the width of the $\rho$ distribution obtained varying all parameters of this category simultaneously. This width is $3 \times 10^{-5}$, which is negligible relative to the systematic error of category (1). The total systematic error is obtained summing those of both categories in quadrature to obtain $1.33 \times 10^{-3}$.

A summary of systematic errors from individual sources (of both categories) is displayed in Table~\ref{tab:SystErrors}. The dominant source of error is the MTAS $\gamma$-tagging efficiency, which defines the order of the systematic error. Though varying spectral characteristics can affect the ad hoc model in the fit, any effects on $\rho$ are minimal as the signal Cu peaks of interest dominate in counts. The statistical error of $1.25\times 10^{-3}$ is of the same order as the overall systematic error. Added in quadrature, these two quantities yield the overall error on $\rho$ of $1.83\times 10^{-3}$.

\begin{table}[ht]
    \centering
    \begin{tabular}{lg} 
      Source   &    \multicolumn{1}{c}{Systematic error}  \\
        \hline\hline
      MTAS $\gamma$-tagging efficiency ($\epsilon$)    &  1.2 \times , 10^{-3}    \\
      K-capture probabilities ($P_K^*,P_K^0$)   &   4.2 \times , 10^{-4}    \\
      Partial \keV{1115} activity ($A\IECs $)   &   2.5 \times , 10^{-4}   \\
      Natural MTAS backgrounds ($\beta\bar{t},\beta$)  &   2.1 \times , 10^{-4}   \\
      SDD-source geometric efficiency ($\eta$)    &  1.8 \times , 10^{-4}\\
      Expected $\gamma$ interaction with SDD ($\eta_\gamma^i$) &    1.3 \times , 10^{-5}   \\
      K fluorescence probability ($\omega_K$)    &   1.8 \times , 10^{-6} \\
        &   \\
      Fit range &    5.2 \times , 10^{-5}    \\
      Binning   &   1.1 \times , 10^{-5}    \\
        \hline\hline
    \end{tabular}
    \caption{Effects of individual sources of systematic error on $\rho$. The statistical error is \RhoResultErrStat . Sources of error stemming from physical limitations (top group) have the associated symbol(s) in parentheses as used in the text. Spectral characteristics (bottom group) have a sub-dominant contribution to the overall systematic error of $1.33\times 10^{-3}$ obtained in the text.}
    \label{tab:SystErrors}
\end{table}

\end{document}